\documentclass[useAMS,usenatbib, fleqn]{mn2e}
\usepackage{color,natbib, amsmath, amssymb, epsfig}
\usepackage{times}

\title{Metallicity bias in the kinematics of the Milky Way stellar halo}
\author[P. R. Kafle, S. Sharma, G. F. Lewis and J. Bland-Hawthorn]
        {P. R. Kafle \thanks{E-mail: p.kafle@physics.usyd.edu.au},
        S. Sharma, G. F. Lewis and J. Bland-Hawthorn \\
	Sydney Institute for Astronomy, School of Physics, A28, 
	The University of Sydney, NSW 2006, Australia}

\begin{document}

\date{Accepted 2013 January 14.  Received 2013 January 13; in original form 2012 December 4}

\pagerange{\pageref{firstpage}--\pageref{lastpage}} \pubyear{2013}

\maketitle
\label{firstpage}

\def \mnras {MNRAS}   \def \apj {ApJ}  \def \apjl {ApJL} 
\def \aj {AJ}         \def \aa {AA}    \def \nat {Nature}
\def \araa {ARA\&A} \def \aapr {A\&AR}
\def \apjs {ApJS}

\def \vl {v_{l}}                   \def \vb {v_{b}}
\def \vr {v_{r}}                   \def \vtheta {v_{\theta}}
\def \vphi {v_{\phi}}              \def \vlossigma {\sigma_{los}}
\def \rsigma {\sigma_{r}}          \def \tsigma {\sigma_{\theta}}
\def \psigma {\sigma_{\phi}}       \def \sigmalos {\sigma_{\rm los}}
\def \vrot {v_{\rm rot}}           \def \vlos {v_{\rm los}}

\begin{abstract}
Here we study the metallicity bias in the radial and tangential velocity 
dispersions, the derived quantity called anisotropy and 
the mean azimuthal velocity profiles of the Milky Way stellar halo 
using Blue Horizontal Branch (BHB) stars taken from {\it SDSS/SEGUE} survey. 
The comparatively metal-rich sample  ([Fe/H] $>-2$) has prograde motion and is found
to have an offset of 40 km s$^{-1}$ in the mean azimuthal velocity with respect to a 
metal-poor sample ([Fe/H] $\leqslant-2$) which has retrograde motion. 
The difference in rotation between the most metal-poor and most metal-rich population
was found to be around 65 km s$^{-1}$. 
For galactocentric distances $r \lesssim 16$ kpc, an offset in velocity 
dispersion profiles and anisotropy can also be seen. 
In the inner regions, the metal-poor population is in average tangential; 
however, anisotropy is found to decrease monotonically with radius independent 
of metallicity. Beyond $r=16$ kpc, both the metal-rich 
and the metal-poor samples are found to have tangential motion. 
The metallicity bias in the kinematics of the halo stars  
qualitatively supports the co-existence of at least two-components in 
the halo having different formation history e.g. in-situ formation 
and formation by accretion.
\end{abstract}

\begin{keywords}
Galaxy: stellar content - Galaxy: halo - Galaxy kinematics: stars: horizontal-branch - stars: abundances
\end{keywords}

\section{Introduction}
The stellar halo is an excellent resource for testing 
theories of galaxy formation \citep{2002ARA&A..40..487F, 2008A&ARv..15..145H}. 
With the advent of new high quality surveys involving hundreds of thousands of stars, 
understanding the formation of the halo has become even more challenging. Under the 
$\Lambda$CDM paradigm of hierarchical formation, the stellar halo is 
thought to have been produced in part by accretion of satellite 
galaxies. Observational evidence for this comes 
from the detection of tidal streams and substructures in the halo 
of both the Milky Way and M31 
\citep[e.g.,][]{1993ARA&A..31..575M, 1995MNRAS.277..781I, 2006ApJ...642L.137B}. 
Quantitative comparison of the amount of substructure and its physical properties seen in 
simulations with that of observations reveals broad 
agreement with the accretion theory 
\citep{2008ApJ...680..295B,2009ApJ...698..567S, 
       2009ApJ...701..776G, 2011ApJ...738...79X,2011ApJ...728..106S}.
Recent cosmological simulations 
\citep[e.g.,][]{2009ApJ...702.1058Z, 2011MNRAS.416.2802F, 2012MNRAS.420.2245M, 2012MNRAS.420..255T} 
including star formation and feedback suggest that the inner regions 
of the halo might be dominated by in-situ component which have
chemical and kinematic properties different from that of the accreted 
component. The kinematic and spatial distributions of the in-situ
component is thought to resemble that of the disk--flattening 
and net prograde rotation. 

Recent observational results also seem to suggest the need for an in-situ component.
In their seminal paper, \citet[hereafter C07]{2007Natur.450.1020C} claim that the inner-halo
component ($r < 15$ kpc) is comparatively metal-rich, has a prograde motion (0-50 km s$^{-1}$), 
with a slightly flattened spatial distribution and with stars on mostly radial orbits. In contrast,
the outer-halo is comparatively metal-poor, has a strong retrograde motion (40-70 km s$^{-1}$),  
has a nearly spherical spatial distribution, with stars on a wide range of orbital eccentricities. 
These results were derived using stars in the Solar neighborhood which were roughly divided into 
inner and outer regions based on the maximum distance from the galactic plane reached by
individual stellar orbits. Their analysis requires the use of proper motions for which 
it is important to have accurate distances. The impact of distance errors on the retrograde signature 
has been debated, e.g. see \cite{2011MNRAS.415.3807S} and \cite{2012ApJ...746...34B}. 
A cleaner distinction between the inner and outer regions 
can only be done by taking a sample of stars which has much wider 
spatial distribution. This was done by \cite{2011MNRAS.411.1480D} using BHB stars  
grouped into four regions in metallicity and distance spaces.  
Their results are in qualitative agreement with C07 and \cite{2010ApJ...712..692C}, 
i.e., the metal-rich population is prograde while 
the metal-poor population is retrograde. 
However, they find that stars between 15-25 kpc are on circular orbits while 
those beyond 25 kpc are on radial orbits. 
The derived quantity called anisotropy \citep{2008gady.book.....B} defined as 
\begin{equation}
\beta= 1 - \frac{\tsigma^{2} + \psigma^{2}}{2\rsigma^{2}}
\end{equation}
describes the nature of the orbits, where $\rsigma$, $\tsigma$, and $\psigma$ are 
radial, polar and azimuthal velocity dispersions respectively.
Negative and positive values of the anisotropy corresponds to the 
tangential and radial dominance of the orbits respectively.
The arbitrary distance ranges make it difficult to compare the 
$\beta$ results of BHB stars with those of C07. 
An arbitrary cut-off in metallicity ([Fe/H] $=-2$) was used to group the stars.  
In our recent work \citep[hereafter K12]{2012ApJ...761...98K}, we
study $\beta$ as a function of radius in greater detail 
and we show that the profile is not a simple monotonic function.
We find that for $r < 15$ kpc, stars are on radial orbits; at larger radii, there
is a sharp drop in $\beta$ with a minimum of
$\beta\approx -1.2$ at 17 kpc; beyond 20 kpc, $\beta$ rises slightly. 
Given the complex variation of $\beta$ with radius, it is important to study the radial 
profile of $\beta$  and $\vrot$ for different metallicities.

This paper is organized as follows. 
In Section 2, we discuss the theoretical aspect of our analysis, 
and the data being used. 
Section 3 presents our result about the metallicity biases in the kinematical 
profiles, $\rsigma(r), \tsigma(r), \psigma(r)$, $\beta(r)$, $\vrot(r)$,  
of the Milky Way stellar halo. In Section 4, we conclude our result
and describe the implications of the biases for the formation of the halo. 

\section{THEORY AND ANALYSIS}\label{sec:theory}
\subsection{Distribution Function and Parameter estimation}
Our main aim is to estimate the rotational velocity and the 
velocity dispersion as a function of radius. 
To do this we model the velocities using the 
Gaussian Velocity Ellipsoidal distribution 
function (GVE DF). A GVE DF with rotation about $z$-axis is given by
\begin{equation} \label{eqn:veDF}
f(r, {\bf v}) = \frac{\rho(r)}{(2\pi)^{3/2}\rsigma\tsigma\psigma} \exp \left[-\frac{1}{2}\left(\frac{v_{r}^{2}}{\sigma_{r}^{2}}+
\frac{v_{\theta}^{2}}{\sigma_{\theta}^{2}}+\frac{(v_{\phi}-\vrot)^{2}}{\sigma_{\phi}^{2}}\right)\right],
\end{equation}
where $\vrot$ is the mean azimuthal velocity.
A GVE DF has been used in the context of the stellar halo to estimate 
the rotation \citep{1980MNRAS.193..295F} and the velocity dispersions
\citep{2004AJ....127..914S, 2009MNRAS.399.1223S, 2012ApJ...761...98K}.
In their studies of halo subdwarf stars,  \citet{2009ApJ...698.1110S} and  
\citet{2010ApJ...716....1B} have found that the tilt of the velocity ellipsoid
is small and consistent with zero, and is therefore ignored in this analysis. 

Ideally, we would like to know the full six-dimensional phase-space information of an 
individual halo star. But presently the proper motion information of the stars in the stellar halo 
beyond the Solar neighborhood ($r \gtrsim 10$ kpc) is not accurate enough to properly constrain 
the tangential motions. We thus resort to the line-of-sight component of the velocity and 
marginalize the DF over the two unknown components of the velocity ($v_{l}$, $v_{b}$).
The marginalized DF can be expressed as,
\begin{equation}
\label{eqn:vlosd}
F(l, b, d, \vlos | \rsigma, \tsigma, \psigma, \vrot) = \iint f(r,{\bf v}) d v_l d v_b.
\end{equation}
Note the marginalization is over the tangential components of the velocities ($v_{l}$, $v_{b}$)
in the heliocentric rest frame, whereas DF (Equation \ref{eqn:veDF}) is expressed in  
the galactocentric frame of reference. For the transformation between these reference frames,
we assume the IAU adopted values of the Sun-Galactic centre to be 8.5 kpc and the 
circular velocity at the Sun's position ($v_{\rm LSR}$) to be 220 km s$^{-1}$. 
We also discuss the effect of changing the value of $v_{\rm LSR}$. 
The values for the peculiar motion of the Sun (U, V, W)$_\odot$ = (+11.1, +12.24, +7.25) in km s$^{-1}$ 
are adopted from \cite{2010MNRAS.403.1829S}. 

Using Equation \ref{eqn:vlosd}, we can express the log-likelihood function as follows 
\begin{equation} 
\label{eqn:likelihood}
\mathcal{L} (\rsigma,\tsigma,\psigma,\vrot) = \sum_{i}^{n} \log F(l_{i}, b_{i}, d_{i}, v_{los_{i}}), 
\end{equation}
where $n$ is the total number of stars. 
Next we use the Markov Chain Monte Carlo (MCMC) to compute the 
posterior distribution of the model parameters.
The median values of the model parameters are quoted as our  
estimates, whereas 16$^{\rm th}$ and 84$^{\rm th}$ percentiles values are used to compute 
the associated uncertainties. 
Note, in the GVE DF (Equation \ref{eqn:veDF}), the density term $\rho(r)$ is not
a function of the model parameters and thus has no effect on the
likelihood analysis. 

%\vspace{-3.mm}
\subsection{Sample}
\begin{figure}
  \includegraphics[width=0.48\textwidth]{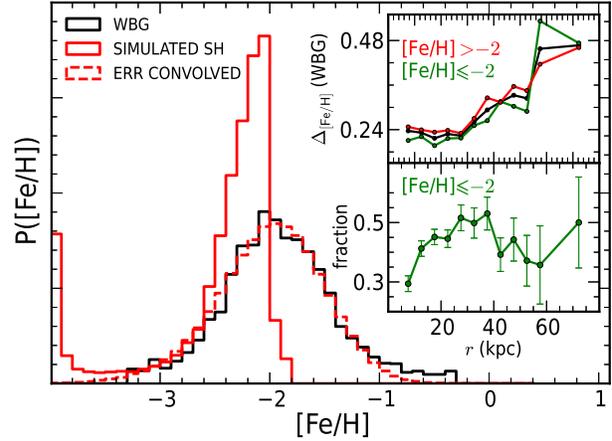}
  \caption{[Fe/H] distribution: Top-inset shows the error in WBG [Fe/H],  
           quoted in the SEGUE database, and the bottom-inset shows the fraction 
           of metal-poor stars with Poisson errors, both as a function of $r$. 
           In the main plot, the black solid line is the [Fe/H] distribution of
           our sample, whereas the red solid line is the [Fe/H] distribution of the simulated 
           stellar halo. The red dashed line is the convolved red line (for [Fe/H]$>-3.5$) 
           with the systematic error of 0.30 and the gaussian random error of 0.40.}
\label{fig:feh}
\end{figure}

BHB stars are luminous and their distances can be measured with
relatively more accuracy compared to other stellar populations,
hence we use them to study the stellar halo.
The sample of BHB stars that we use is taken from \citet[X11]{2011ApJ...738...79X}
comprising 4985 BHB stars obtained from {\it SDSS/SEGUE}.
To minimize contamination from disk stars, 
we select stars that are well removed from the plane of the disk ($|z|>4$ kpc). 
We do not impose any kinematic limits to obtain our sample and  
hence for the purpose of kinematic studies this sample of BHB stars
can be considered to be unbiased.
To improve the distances, we recalibrate X11 distances using a color-magnitude 
relation from \citet{2011MNRAS.416.2903D}. 
The dispersion in $g$-band magnitudes of the data results a distance uncertainty of 6 per cent. 
In the radial velocity measurements, 95 per cent of our sample has an uncertainty of less than 9 km s$^{-1}$. 
The final sample size of the BHB stars we use is 4386.

The X11 catalog of BHB stars originally did not provide the corresponding metallicity values.
To obtain [Fe/H], we thus cross-check SEGUE database for each star against the X11 sample. 
The [Fe/H] metallicity we adopt are the estimates based on \cite{1999AJ....117.2308W}
which is provided in the SEGUE database under the heading `feh\_WBG'. 
In Figure \ref{fig:feh}, we present the [Fe/H] distribution of our sample. 
In the inset of the figure, the errors in WBG [Fe/H] estimates are shown 
as a function of distance ($r$) for both the metal-rich (red line) and metal-poor (green line)
sub-sample as well as for the total sample (black line).
Note, errors in WBG [Fe/H] beyond $r=35$ kpc are significantly larger ($>0.27$ dex).
Also shown in the figure as a red line is the metallicity of BHB
stars in the simulated stellar halos by \cite{2005ApJ...635..931B} sampled using the code
Galaxia \citep{2011ApJ...730....3S}. 
The code Galaxia takes the N-body particles and their given
age and metallicity distributions and then uses Padova stellar isochrones
to generate stars.
A sharp upper cut-off in the metallicity at around [Fe/H] $\approx$ -1.7 can be
seen and this constraint comes purely from the theory of stellar
evolution. According to the synthetic isochrones 
used by Galaxia (Padova isochrones), there are no BHB stars 
with age less than 13 Gyr and metallicity greater than -1.7.
In reality, globular clusters and the field stars \citep{2003ApJS..149..101B} are found to have
metal-rich BHB stars.
We find that when the simulated BHB stars are convolved with an 
uncertainty of 0.40 dex and an offset of 0.30 dex we can match the 
metallicity of the SEGUE BHB stars (dashed red line).
The disagreement among the theoretical prediction and the observed metallicities
of the BHB stars could possibly be either because the synthetic 
isochrone do not correctly model the BHB stars or may be because
of inaccuracies in the computation of WBG BHB metallicities quoted in SEGUE database.

Also shown in Figure \ref{fig:feh} is the fraction of metal-poor stars
([Fe/H] $\leqslant-2$) as a function of radius. It can be seen that the fraction increases with
distance suggesting that the inner regions are slightly biased towards being metal 
enriched. However, in the outer parts, it starts to fall again due to the sharp increase 
in uncertainty of [Fe/H] at larger distances, 
e.g., the outer part could be metal-poor but a large 
uncertainty in [Fe/H] will spuriously classify metal-poor stars 
as relatively metal enriched.

\section{Results: Metallicity Biases in Kinematics}\label{sec:kinematics}

We present here our estimates of the model parameters $\rsigma(r),\tsigma(r),\psigma(r)$, 
$\beta(r)$ and $\vrot(r)$ in comparatively metal-rich 
([Fe/H] $>$ -2) and metal-poor ([Fe/H]$\leqslant$ -2) bins.
To compute the kinematic profiles as a function of radius we choose a 
moving centre scheme for binning.
Briefly, in this binning scheme a set of equi-spaced positions in $r$
are chosen and then at each position an equal number of points ($n_{\rm bin}$) 
from the central value are used to estimate the velocity dispersions.
The number density of stars in $r$ is not uniform and hence we report the
mean $r$ of the points in each bin as its final position.
It should be noted that the bins are overlapping in this scheme;  
a detail discussion on associated pros and cons of using the 
moving centre bins are given in K12. 

\subsection{Bias in Mean Azimuthal Velocity Profile}\label{sec:vrot}

\begin{figure}
    \includegraphics[width=0.475\textwidth]{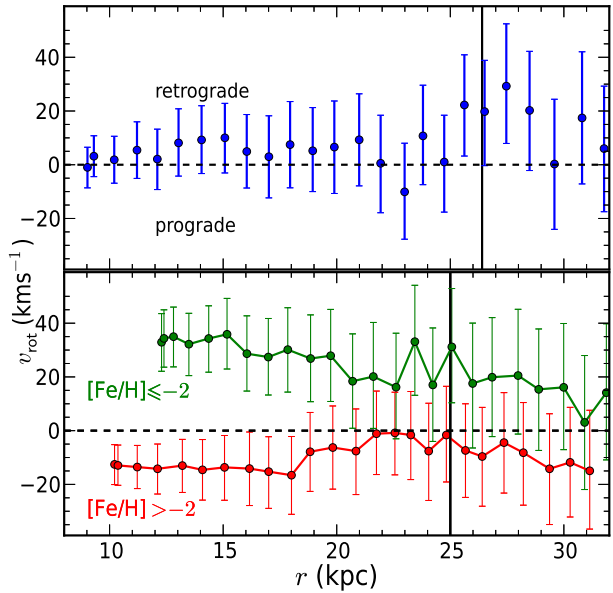}
  \caption{Metallicity bias in $\vrot$: $\vrot$ estimates in two 
           metallicity bins [Fe/H] $>-2$ (red) and [Fe/H] $\leqslant-2$ (green). 
           The $\vrot$ profile for the total sample is shown with the blue markers.
           Black vertical solid line demarcates bins that have stars with $r<30$ kpc.}
\label{fig:vrot}
\end{figure}

\begin{figure}
    \includegraphics[width=0.475\textwidth]{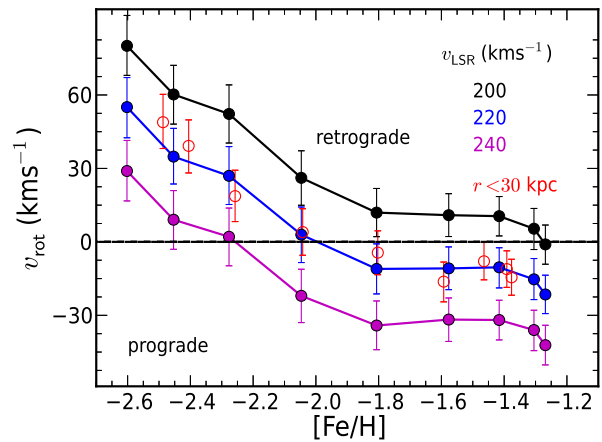}
  \caption{$\vrot$ dependence on [Fe/H] and the effect of $v_{\rm LSR}$: 
           The blue dots with the error bars show the $\vrot$ as a function of  
           [Fe/H] for the sample of BHB stars,
           with $v_{\rm LSR}$ = 220 km s$^{-1}$, whereas red open circles 
           are the values for the sample restricted to $r<30$ kpc.
           Black and magenta markers with the error bars are the corresponding estimates 
           for $v_{\rm LSR}$ = 200, 240 km s$^{-1}$ respectively.}
\label{fig:vrotmetal}
\end{figure}

We first compute the profile of mean azimuthal velocity 
($\vrot$ in Equation \ref{eqn:veDF}) with $n_{\rm bin} = 600$.
The blue markers in Figure \ref{fig:vrot} show the $\vrot$ profiles of 
the total sample of BHB stars. 
It can be seen that the $\vrot(r)$ profile of the overall population is nearly constant. 
The bottom panel shows the profiles 
for the metal-rich and metal-poor sample separately.
In the inner region, $r<18$ kpc, the metal-rich population 
is found to have a prograde motion of $\vrot$ = 10 km s$^{-1}$ which
declines slightly in the outer-region and attains null rotation at $r> 20$ kpc.
In contrast, the metal-poor population has a substantial retrograde motion of 
30 km s$^{-1}$ for $r < 18$ kpc. 
At larger $r$, $\vrot$ values of the metal-poor population also decrease to about 18 km s$^{-1}$.
Figure \ref{fig:vrot} is restricted to $r<32$ kpc as beyond here
the large uncertainty in [Fe/H] (see Figure \ref{fig:feh}) washes out the distinction between the metal-rich and 
the metal-poor population and makes the $\vrot$ profiles
to converge to a common value. 
Additionally, the error in $\vrot$ also begins to increase with $r$.
There are no estimates for metal-poor samples for $r<12$ kpc
which is due to the fewer metal-poor stars in our sample in that regime 
than the metal-rich stars, as expected.
   
\cite{2011MNRAS.411.1480D} also derive $\vrot$ and 
find that in a large bin ranging $10<$ $r$ / kpc $<50$, 
the metal-rich population has a net prograde motion of 15 km s$^{-1}$ 
whereas the metal-poor population has a net retrograde motion of 25 km s$^{-1}$.
In agreement with Figure \ref{fig:vrot}, it can be said that their estimates
of $\vrot$ are more likely the better representation of the inner halo within $r<18$ kpc,
but definitely not of the halo out to $r=50$ kpc. 
Note that their sample of BHB stars are not selected based on Balmer line-strengths,
and thus we anticipate significant contamination  
from the Blue Straggler population as well as 
from the cooler and low gravity candidates of the horizontal branch 
stars \citep{2004AJ....127..899S} such as RR Lyrae.
The fact that the metal-rich and metal-poor profiles do not have a
very strong dependence on radius means that we can neglect the radial 
variation of $\vrot$ and study the $\vrot$ as a function of metallicity;  
this is shown in Figure \ref{fig:vrotmetal}. 
A switch from retrograde to prograde motion at [Fe/H] $\sim$ -1.8
is notable in this figure, suggesting a separate metal-rich
component. 
We checked that this transition also remains for the sub-sample of
stars within $r<30$ kpc, for which errors in [Fe/H] is $<0.27$ dex. 
Given about 0.22 dex uncertainity in [Fe/H],  
we find that a two component theory, $\vrot$ being a step function of [Fe/H], 
can explain the above profile, but a smooth transition to the low 
metallicity end cannot be ruled out. 
From the fact that the $\vrot$ profile is not symmetric 
about [Fe/H] $=$ -2, we can infer that the actual transition to the
low metallicity component will be at [Fe/H] $<$ -2. 
This makes sense in the context of the metallicity distribution 
of BHB stars for the simulated stellar halo that was shown
in Figure \ref{fig:feh}, 
namely that the synthetic isochrones predict very few BHB stars with 
[Fe/H] $>$ -2.

The $v_{\rm LSR}$ has a correlation with $\vrot$ estimates. 
To investigate this we repeat the analysis for different values of 
$v_{\rm LSR}$ and the results are shown in Figure \ref{fig:vrotmetal}. 
The effect of changing $v_{\rm LSR}$ is to shift the curves up or down by the 
corresponding amount but keeping the feature of the profile nearly intact. 
Hence if $v_{\rm LSR}$ is $\sim$250 km s$^{-1}$
(for $R_{\odot} = 8.5$ kpc; \citealt{2010MNRAS.402..934M}),
then the metal-poor population will have nearly zero rotation and the metal-rich population will 
have a strong prograde rotation.
This is consistent with the picture that the metal-poor stars that dominate the outer regions of the 
halo is formed by accretion of multiple satellites without any preferred 
direction of rotation. On the other hand, the metal-rich stars   
that dominate the inner-halo could have their origin related to the disc 
through the in-situ stars, or due to a massive satellite which
happened to have the rotation in the same direction as disc. 
However, irrespective of the adopted value of $v_{\rm LSR}$, 
the difference in the rotational properties of metal-rich and
metal-poor population remains intact and our results show that 
the difference in $\vrot$ between the most metal-rich and most metal-poor is at least 65 
km s$^{-1}$.

\subsection{Bias in Velocity Dispersion Profile}\label{sec:veldis}

\begin{figure}
    \includegraphics[width=0.475\textwidth]{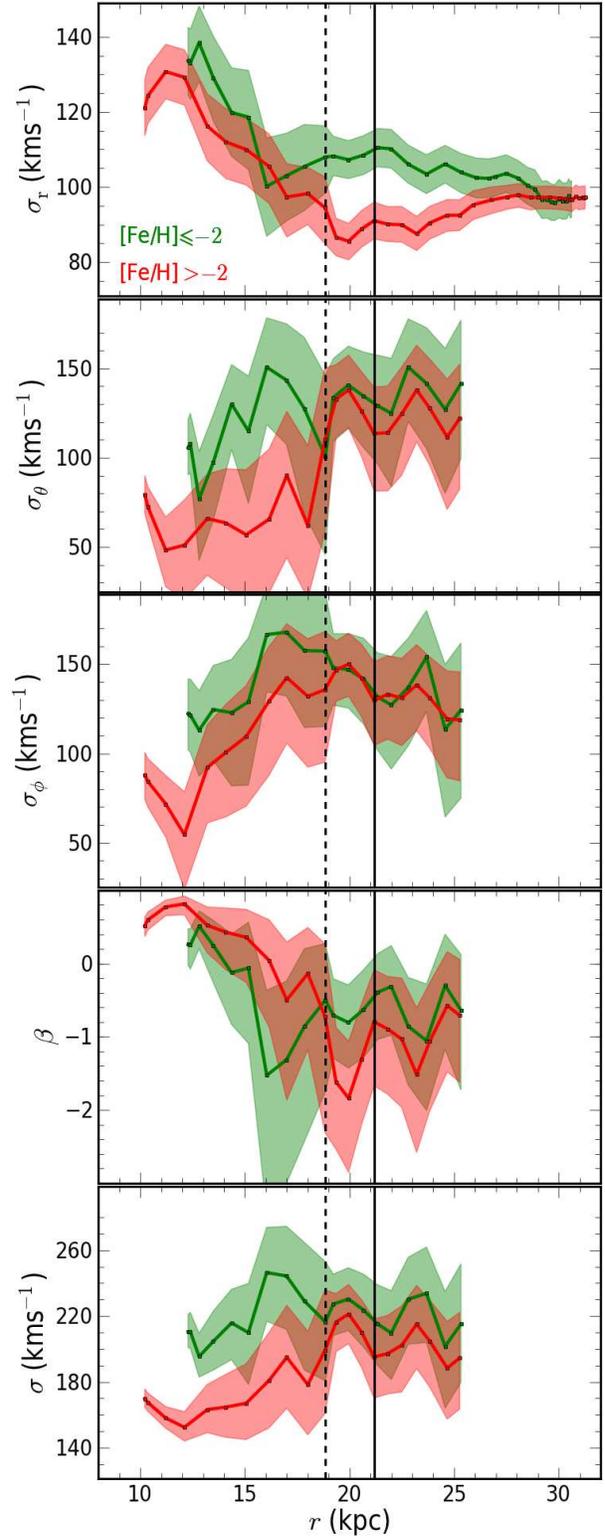}
  \caption{Metallicity bias in the velocity dispersions: 
           red and green lines are the estimated values of velocity dispersions 
           and corresponding $\beta$ and  $\sigma$ in the comparatively metal-rich ([Fe/H] $>-2$) 
           and metal-poor ([Fe/H] $\leqslant-2$) bins. 
           Shaded area denote the errors associated.}
\label{fig:veldis}
\end{figure}

Apart from $\vrot$, do the metal-rich and metal-poor subsamples differ in 
any other kinematic property? 
To answer this, we study the metallicity dependence of
$\rsigma(r),\tsigma(r),\psigma(r)$ and $\beta(r)$, and are shown in Figure \ref{fig:veldis}.
The dispersions are measured in radials bins with $n_{\rm bin} = 600$
stars for $r<18.9$ kpc and $n_{\rm bin} = 1200$ for $r>18.9$ kpc. 
This demarcation is shown by the vertical dashed line. 
We have to use larger $n_{\rm bin}$ in the outer parts because
fewer stars do not constrain the $\tsigma$ and $\psigma$. 
This is because in the outer parts the radial 
velocity carries less information about the tangential components.
A drawback of $n_{\rm bin}$ being large is that the spatial resolution is compromised. 
Nevertheless, large bins are useful to study general trends. 
In K12 it was shown that the overall trend in dispersion and 
$\beta$ profile of BHB stars 
remains intact even as $n_{\rm bin}$ is increased from 750 to 1200,
except for the fact that the profiles with large $n_{\rm bin}$ are
smoothed versions of those with lower $n_{\rm bin}$. 

From the figure we can see that below $r=16$ kpc the trends 
in all the five listed profiles are similar for both the meal-rich and 
metal-poor populations. Namely, with increase in $r$, the $\rsigma$
decreases, the $\tsigma$ and $\psigma$ increase and the $\beta$
decreases. However, the profiles are offset from each other.
Also, the total dispersion $\sigma$ ($= \sqrt(\rsigma^2 + \tsigma^2 + \psigma^2$))
is found to have an offset (40 km s$^{-1}$).
Overall the metal-poor population has lower $\beta$ than metal-rich. 
The maximum difference is at $r=16$ kpc, with $\tsigma$ being very high 
and the $\beta$ being very low for the metal-poor population.
Going from $r=16$ to $r=20$ kpc, one finds that the metal-poor population 
has almost constant $\rsigma$ whereas for the metal-rich population 
the $\rsigma$ drops. 
The $\beta$ profile for the metal-poor population remains relatively 
constant whereas for the metal-rich it keeps on decreasing with 
minimum value being about -2. 

Beyond $r=20$ kpc, the metal-rich and metal-poor profiles slowly 
start to converge. This is probably related to the fact that 
the uncertainty in [Fe/H] increases sharply beyond $r=30$ kpc. 
So any kinematic differences between the metal-poor and metal-rich
are expected to be washed out if stars beyond $r=30$ kpc are
included. Note in our scheme the bins are overlapping, so a bin 
at a given radius can contain stars within a wide range in $r$. 
To show this more clearly, in the figure we plot a solid vertical 
line that demarcates bins that have stars with 
$r<30$ kpc. It can be seen that as we move to the right of this line 
the two $\rsigma$ profiles start to converge. 

The fall at $r=16$ kpc in $\rsigma$ of the metal-rich population 
roughly coincides with the break radius in the density distribution of the halo stars; 
this fall  was also noticed by \cite{2011MNRAS.411.1480D} in the $\sigma_{\rm los}$ profile.
In a recent paper \citep{2013ApJ...763..113D}, it was postulated that 
a break radius in the density profile of a halo could be due
to accretion of a massive metal-rich accretion event and  
a signature of this would be the fall in $\rsigma$ of the 
metal-rich population. 
We do see this dip in the metal-rich BHB stars but they are found 
to be on tangential orbits while shell-like structures in the \cite{2005ApJ...635..931B}
simulations are generally found for satellites on highly radial orbits.
It remains to be seen if a satellite on more milder radial orbit 
can still give rise to a break in density.

%+++++++++++++++++ 5<r/kpc<20++++++++++++++++++++++++++++++++++++++++++++++
\begin{figure}
    \includegraphics[width=0.475\textwidth]{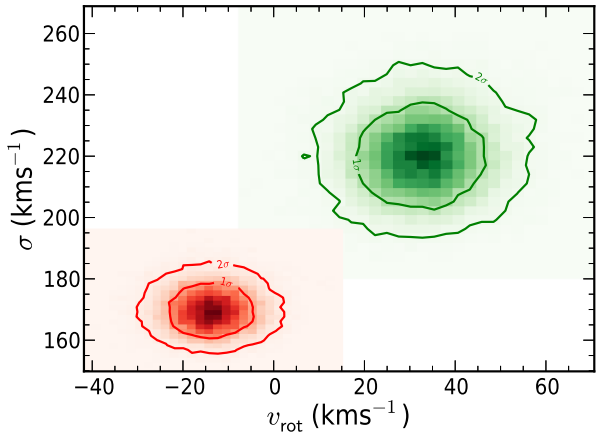}
  \caption{Metallicity bias for the sample in a range $5<r$/kpc$<20$ : 
           the $1\sigma$ and $2\sigma$ confidence contours for the 
           joint-probability distributions of the parameters in
           two metallicity bins [Fe/H] $>-2$ (red) and 
           [Fe/H] $\leqslant-2$ (green) respectively.}
\label{fig:contours}
\end{figure}

\begin{table}
\centering
\caption{Estimated parameters in the comparatively metal-rich and metal-poor bins
         in the range $5<r$/kpc$<20$ for two different cases namely, 
         when Sagittarius stream and Virgo-overdense region masked and unmasked.}
\begin{tabular}{c|cccc}
\hline
\hline
Saggi-Virgo& [Fe/H]         &$\beta$ & $\sigma$ (kms$^{-1}$) & $v_{\rm rot}$  (kms$^{-1}$)\\
\hline
Unmasked&$>-2$          & $0.3^{+0.1}_{-0.2} $ & $ 170^{+6}_{-6}$  & $-14^{+6}_{-6}$  \\
Masked&$>-2$          & $ 0.3^{+0.1}_{-0.2}$ & $ 170^{+6}_{-6}$  & $-15^{+6}_{-7} $  \\
Unmasked&$\leqslant-2$  & $-0.4^{+0.3}_{-0.4}$ & $220^{+11}_{-11}$ & $33^{+9}_{-10}$   \\
Masked &$\leqslant-2$  & $-0.4^{+0.3}_{-0.5}$ & $220^{+12}_{-11}$ & $35^{+10}_{-10}$ \\
\hline
\end{tabular}
\label{table:r5_20} 
\end{table}

It is apparent in the Figures \ref{fig:vrot} and \ref{fig:veldis}
that the analyses are done in smaller bin sizes and consequently, the errors
for each bin are larger. 
Thus, here we also provide additional measures of our model parameters,
$\sigma$, $\beta$ and $\vrot$, for the metal-rich and the metal-poor samples 
covering the larger radial range $5<r$/kpc$<20$.
In Figure \ref{fig:contours}, we show the joint probability 
distributions of the model parameters for these sub-samples, 
where for the metal-poor and the metal-rich samples
the confidence contours ($1\sigma$ and $2\sigma$) are shown 
in green and red colors respectively.
The estimated values of these parameters are given in Table \ref{table:r5_20}.
The differences among the estimated values of the kinematic parameters 
$\sigma$ and $\vrot$ of the metal-rich and the metal-poor samples
are found to be $\sim50$ km s$^{-1}$.
Also given in Table \ref{table:r5_20} are the estimates of the parameters 
for the sample masked to remove the contamination from 
Sagittarius stellar stream and the Virgo overdensity .
The cuts chosen to mask these structures are purely geometrical
and are same as given in Section 3.4 of K12 
(for the further details about the cuts see the references provided there within).
It can be seen in the Table that the exclusions of these two substructures 
has negligible effect on the final results.

%+++++++++++++++++ In a bin++++++++++++++++++++++++++++++++++++++++++++++++
\section{CONCLUSIONS}
In this paper, we derived the mean azimuthal velocity,
the velocity dispersion and the anisotropy profiles of the 
Milky Way stellar halo, using BHB stars, as a function
of distance $r$ and metallicity [Fe/H].

The comparatively metal-rich ([Fe/H] $>$ -2) and 
metal-poor ([Fe/H] $\leqslant$ -2) populations of BHB stars 
within $r<16$ kpc have distinct kinematic properties with offsets 
of 40 km s$^{-1}$ in both $\vrot$ and $\sigma$. 
Additionally, an offset is also found in $\rsigma$,
$\tsigma$, $\psigma$ and $\beta$. This suggests 
the presence of at least two sub-populations in the halo.
When $\vrot$ is studied as a function of [Fe/H], we find 
that $\vrot$ decreases monotonically and then saturates to a
 constant value for [Fe/H] $>-1.8$. The saturation again suggests the 
existence of a distinct metal-rich component. The difference in
rotation between the most metal-poor and most metal-rich population
was found to be around 65 km s$^{-1}$. 

Overall the $\beta$ profile for both the metal-rich and the metal-poor populations
show similar trends with radius, i.e., $\beta$ decreases with radius 
and then saturates at a negative value. This is in agreement 
with \cite{2011MNRAS.411.1480D} in the sense that $\beta$ is more a function 
of radius than metallicity, but differs slightly from C07 who suggest that the metal-rich 
halo is radial while the metal-poor has a wide range of values of
$\beta$. Our results suggest that below $r<15$ kpc, the metal-rich population
has radial orbits and the metal-poor population is on average tangential. But 
in the range $15<$ $r$/kpc $<25$, both the metal-rich and metal-poor populations are
predominantly tangential.  Finally, regarding the spatial distribution of the metallicity  
we find that although the inner regions ($r<15$ kpc) are slightly biased 
towards being metal-rich, in general the metal-rich and metal-poor 
populations have substantial spatial overlap.

There are two factors which make it difficult to compare the 
BHB results with the findings of C07. Firstly, BHB stars are 
in general metal-poor so they preferentially sample the metal-poor 
population of the halo. Secondly, there is significant 
uncertainty  in estimated metallicity. The uncertainty in fact  
increases sharply beyond 30 kpc. The above two factors 
will lead to significant overlap between the metal-poor
and metal-rich populations and will smooth out the differences between 
the two populations. We do find evidence of this in the outer 
parts ($r>25$ kpc) where a convergence in $\rsigma$ and $\vrot$ is noticed
for the distinct populations due to the larger measurement uncertainties.
Thus the physical differences between the two populations could be 
even stronger, but a deeper stellar survey would be required to confirm this.

Note a contamination of the BHB stars with other populations 
still a possibility. We find that theoretical isochrones do not 
predict BHB stars with [Fe/H] $>-1.7$ and have ages less than 13 Gyr. 
A distribution of synthetic BHB stars can only be matched 
if we convolve with an uncertainty of 0.40 dex and allow for 
an offset of 0.30 dex with respect to the metallicity of the 
SEGUE BHB stars. 
However, some probable field BHB stars (e.g. BD+$49^{\circ}$2137 from \citet{2003ApJS..149..101B}) 
have been found to be metal-rich, thus it could be true that 
the synthetic isochrone do not model BHB stars well.
Within $r=$16-20 kpc, we find that the $\rsigma$ and $\beta$ of 
the metal-rich population decrease sharply while that of the metal-poor population 
show a slight increase. The fall in $\sigma$ of the metal-rich population 
roughly coincides with the break radius in the density distribution of the halo stars. 
As suggested by \cite{2013ApJ...763..113D} this could be a signature of accretion of a 
massive metal-rich accretion event.

To conclude, we find that the metal-rich and the metal-poor 
populations of the halo BHB stars have distinct differences in 
kinematic properties. 
But note that there is no sharp transition from one population
to another e.g. C07 [Fe/H] distributions of 
the inner-halo and the outer-halo peak at -1.6 and -2.2 respectively but
have a spread which are larger than the difference of their peak values. 
Therefore a simple division into two groups by [Fe/H]
will include substantial numbers of the outer-halo in the metal-rich 
group and vice-versa.

The difference in the metal-rich and metal-poor seen by us is in 
qualitative agreement with earlier observational findings of 
C07, \cite{2010ApJ...712..692C}, and \cite{2011MNRAS.411.1480D} and 
supports the dual formation scenario as seen in cosmological 
simulations, i.e., the inner halo is dominated 
by an in-situ component and the outer halo by an accreted component 
\citep[e.g.,][]{2009ApJ...702.1058Z, 2011MNRAS.416.2802F, 2012MNRAS.420.2245M}. 
However, to accept or reject the above hypothesis, a detailed 
comparison of observations with simulations still needs to be done.
Also a comparative study of the kinematics of stellar 
populations other than BHB stars such as 
turn-off stars (fairer sample in metallicity) or giants (probing farther 
region of the outer-halo) or standard candles
e.g. RR Lyrae (accurate distances) is needed. 

\section*{ACKNOWLEDGEMENTS}
P.R.K acknowledges University of Sydney International Scholarship for the support of his candidature. 
G.F.L acknowledges support from ARC Discovery Project (DP110100678) and Future Fellowship (FT100100268). 
J.B.H. is funded through a Federation Fellowship from the ARC and 
S.S. is funded through ARC DP grant 0988751 which supports the HERMES project.
We sincerely thank the anonymous referee for comments that helped to improve the paper.
\bibliographystyle{mn2e}
\bibliography{ms}

\label{lastpage}
\end{document}